\documentclass[aps,prl,twocolumn,showpacs,superscriptaddress,groupedaddress]{revtex4}
\usepackage{graphicx}
\usepackage{epstopdf}
\usepackage{tabularx}
\usepackage{amsmath}
\usepackage{array}
\usepackage{natbib}
\usepackage{color}
\usepackage{romannum}
\usepackage[Symbol]{upgreek}

\begin{document}

\title{Type-\Romannum{2} Zeeman slowing: Characterization and comparison  to conventional radiative beam slowing schemes}

\author{M. Petzold, P. Kaebert, P. Gersema, T. Poll, N. Reinhardt, M. Siercke and S. Ospelkaus}
\email[]{petzold@iqo.uni-hannover.de}
\homepage[]{www.iqo.uni-hannover.de}
\affiliation{Institute for Quantum Optics, University of Hanover}

\date{\today}

\begin{abstract}
We describe a novel Zeeman slowing method reported in (\citet{petzold_zeeman_2018})  and compare it to conventional radiative beam slowing schemes.
The scheme is designed to work on a type-\Romannum{2}  level structure making it particularly attractive for radiative beam slowing of molecules. Working on the D$_{1}$-line of atomic $^{39}$K, we demonstrate efficient slowing of an atomic beam from $\mathrm{400 \, m \, s^{-1}}$ down to $\mathrm{35 \, m \, s^{-1}}$ with a final flux of  $3.3 \cdot 10^{9} \, \mathrm{cm}^{-2}\mathrm{s^{-1}}$. We give experimental details and compare our results to other established radiative slowing schemes in atomic and molecular physics. We  find type-\Romannum{2}  Zeeman slowing to outperform white-light slowing commonly used in molecular beam slowing and to be comparably efficient as traditional type-\Romannum{1} Zeeman slowing being the standard beam slowing technique in atomic physics.

\end{abstract}


\maketitle


\section{Introduction}

The ability to routinely cool atoms to ultracold temperature has revolutionized atomic physics. Laser cooling led to the production of quantum degenerate gases of bosons and fermions which nowadays are used in many applications ranging from precision measurements to quantum simulation of interacting many body systems.
In these experiments, a large range of exciting research opportunities would open up by using molecules instead of atoms \cite{carr_cold_2009, bohn_cold_2017}. Molecules are ideal candidates for the search for a permanent electric dipole moment of the electron \cite{baron_order_2014, hudson_improved_2011} and  the long-range and anisotropic interactions between polar molecules allow for the study of dipolar quantum many-body systems \cite{baranov_condensed_2012}.

 Unfortunately, the cooling process for molecules itself challenges experimentalists due to the complex internal level structure exhibiting rotational and vibrational degrees of freedom. Nevertheless, there are a variety of approaches, which led to the production of ultracold molecular ensembles. Temperatures below $1\,\mathrm{\upmu K}$ have been achieved for a very specific class of diatomic molecules that can be assembled from ultracold alkali atoms \cite{ospelkaus_quantum-state_2010}. Polyatomic molecules have been captured in electrostatic traps and have been cooled down to $400 \, \mathrm{\upmu K}$ by Sisyphus-cooling \cite{prehn_optoelectrical_2016}. 
 
 Recently, laser cooling of molecules with highly diagonal Franck-Condon structure has developed at a tremendous pace: Initial demonstration of transverse cooling of a molecular beam \cite{shuman_laser_2010} has soon been followed by the demonstration of  molecular magneto-optical traps \cite{barry_magneto-optical_2014} and  sub-Doppler cooling of molecules to temperatures in the  $\mathrm{\upmu K}$ range \cite{truppe_molecules_2017, cheuk_lambda-enhanced_2018}. Furthermore, it has been pointed out that there seem to be promising ways to further cool the molecules of interest  by evaporative cooling down to quantum degeneracy \cite{quemener_shielding_2016-1, gonzalez-martinez_adimensional_2017}. However, at the current stage, laser cooling experiments of and with molecules suffer from low molecule numbers, the main limitation already resulting from inefficient molecular beam slowing to velocities trappable by magneto-optical traps.  Typically, experiments start with molecular beams from cryogenic buffer gas cells with velocities in the $90-150 \, \mathrm{m \, s^{-1}}$ range \cite{hutzler_buffer_2012}. The molecular beams are then being slowed down by radiative slowing techniques such as white light slowing \cite{barry_laser_2012, hemmerling_laser_2016} or chirped light slowing \cite{truppe_intense_2017,yeo_rotational_2015}. 
 
In \cite{petzold_zeeman_2018} we have proposed a new radiative slowing method for molecules, which should overcome many of the difficulties in slowing  molecular beams to low velocities trappable in magneto-optical traps. The idea is to start from well-known and tremendously efficient Zeeman slowing of atoms \cite{phillips_laser_1982} (which in its standard form fails for molecules due to the molecules' complex level structure) and find ways and tricks to make the scheme applicable for molecules working on a  type-\Romannum{2} ($\mathrm{J}\rightarrow \mathrm{J'}=\mathrm{J-1}$ )
level transition.  We call the resulting scheme a  type-\Romannum{2} Zeeman slower in contrast to the traditional atomic (type-\Romannum{1}) Zeeman slowing working on a closed $J$ to $J+1$ transition. 

Here we discuss in  detail  type-\Romannum{2} Zeeman slowing on the D$_{1}$-line of $^{39}$K. We start by a qualitative comparison of different radiative beam slowing techniques. We then  present a test experiment in detail, which was briefly mentioned in \cite{petzold_zeeman_2018}. 
 We describe our experimental apparatus and present the results of type-\Romannum{2} Zeeman slowing. Finally, we compare our results to  type-\Romannum{1} Zeeman slowing as a benchmark for an efficient radiative slowing scheme for atoms and white light slowing as a radiative molecular beam slowing technique. 

\section{Characteristics of different radiation pressure beam slowing methods}

\begin{center}

\begin{figure*}
\onecolumngrid

	\includegraphics[width=0.9\textwidth]{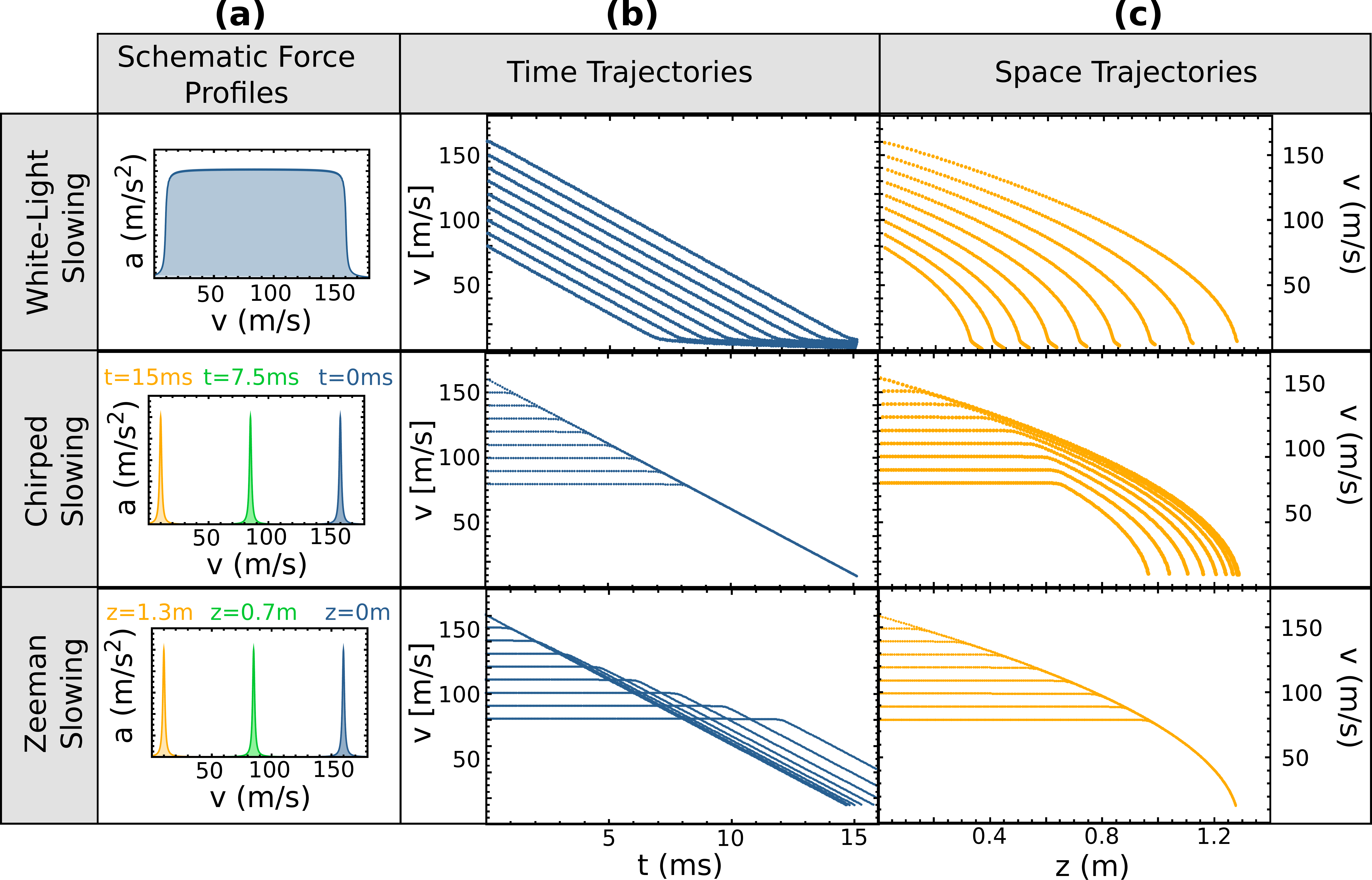}%

	\caption{\textbf{(a)} Application of the deceleration force when $\mathrm{a_{max}=10^{5}\,m\,s^{-2}}$ of different radiation pressure slowing methods while decelerating particles from $\mathrm{160\,m\,s^{-1}}$ down to $\mathrm{10\,m\,s^{-1}}$. \textbf{(b)} One dimensional velocity trajectory simulations over time t during the slowing process with initial velocities ranging from $\mathrm{160\,m\,s^{-1}}$ to $\mathrm{80\,m\,s^{-1}}$. \textbf{(c)} One dimensional velocity trajectory simulations over slowing distance z during the slowing process.	\label{SlowingMethods} Note that Zeeman slowing is the only method where all particles reach the target velocity at the same point in space. Also note that for chirped slowing and Zeeman slowing the deceleration stops at the target velocity, whereas for white-light slowing the remaining finite deceleration force leads to further deceleration.}
\end{figure*}
\end{center}

In order to be trapped in magneto-optical traps the particles coming from a cold molecular beam source need to be slowed from their initial velocities of a few hundred $\mathrm{m\,s^{-1}}$ down to a few $\mathrm{m\,s^{-1}}$.  This can be achieved by shining in a resonant laser beam counterpropagating to the molecular beam. By directed absorption and undirected emission of photons the particles are slowed down by the radiation pressure. The frequency of the slowing laser dictates the velocity classes being slowed down due to the corresponding Doppler shifts.For efficent deceleration the resonance condition has to be met by the slowing laser over the whole velocity range. There are three different methods, which are currently used to accomplish this. Firstly, the slowing laser can be frequency broadened so that all velocity classes are resonant (white-light slowing). Secondly, the frequency of the slowing laser can be chirped in time so that the resonance condition is fulfilled while the particles are slowed down (chirped light slowing). Lastly, the resonance frequency of the particles is altered in a spatially inhomogeneous magnetic field (Zeeman slowing). Figure \ref{SlowingMethods} \textbf{(a)} shows the force profiles of these three different slowing methods along with corresponding time \textbf{(b)} and space \textbf{(c)} trajectories of particles with different velocities during the slowing process.

The bottom row shows the corresponding graphs for Zeeman slowing, where the force profile is altered along the slowing path.
It is the only method where all particles reach their final velocity at the same point in space.
Whereas for white-light slowing and chirped light slowing the particles reach the target velocity at different points in space and  the width of the velocity distribution is thereby projected on the slowing path. Zeeman slowing and white-light light slowing are continuous methods while chirped light slowing, due to its nature will lead to pulses of slow particles.


When it comes to trap loading the most often used technique in atomic experiments is Zeeman slowing. The reason lies in the fact, that for trap loading the particles have to traverse a specific trap volume with a velocity equal to or lower than the corresponding capture velocity of the trap. So far we have only talked about one dimension. Any additional transverse velocity can lead to a loss of particles due to transverse spreading.  This effect is minimized if the time the particles need to reach the trap volume is reduced. This is another benefit of the Zeeman slowing technique as all particles first fly with their initial velocity and are only slowed down when it is needed, therefore losses due to transverse spreading are efficiently lessened. Furthermore, it combines the advantageous  characteristics of the other slowing methods, being continuous as in white-light slowing and having a well defined final velocity as chirped light slowing. All these effects add up and make Zeeman slowing the most effective radiative slowing method for the purpose of trap loading.

\section{Type-\Romannum{2} Zeeman slowing}

\begin{center}
\begin{figure*}
\onecolumngrid

	\includegraphics[width=0.9\textwidth]{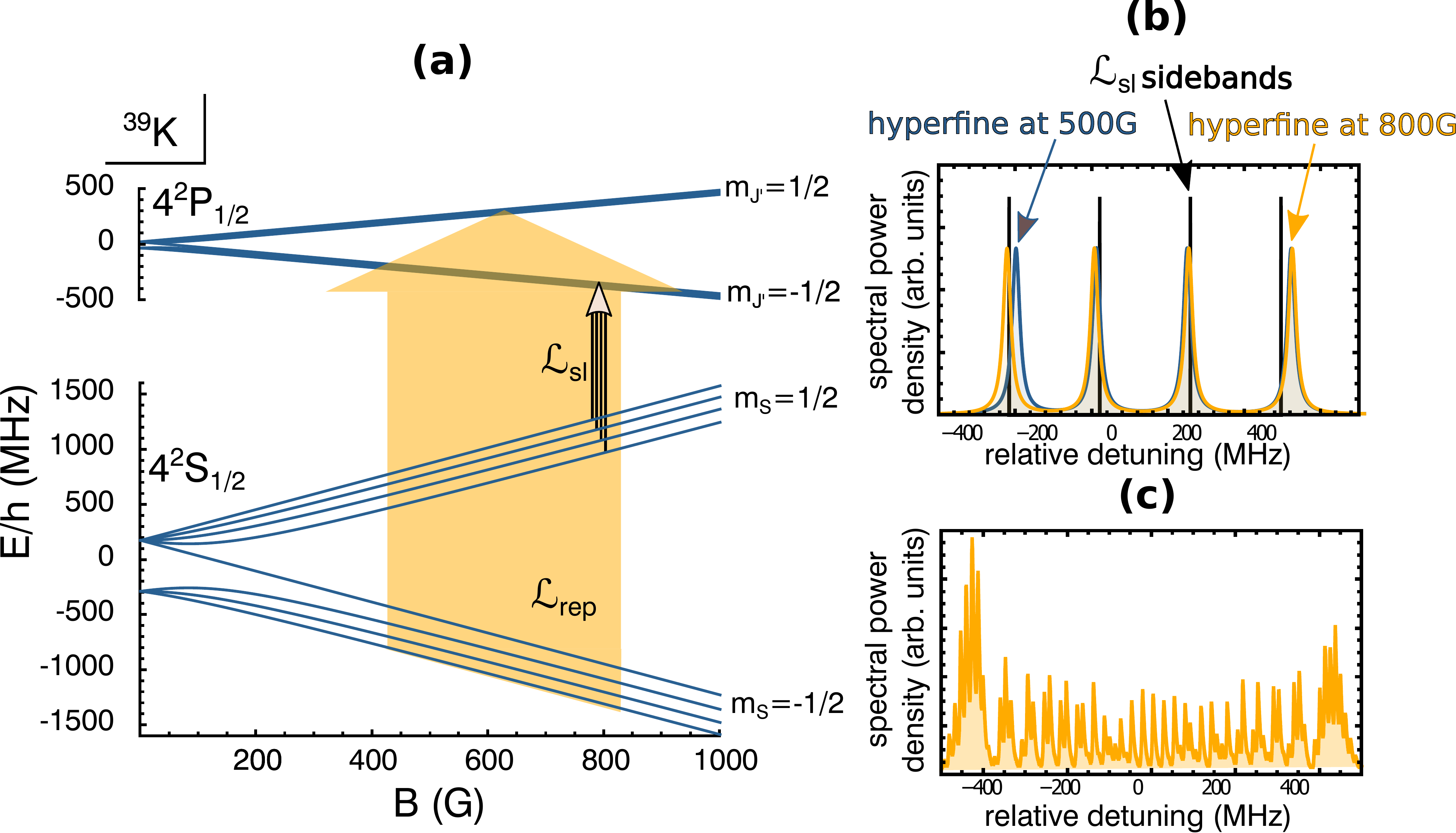}%

	\caption{\textbf{(a)} Type-\Romannum{2} Zeeman Slower scheme for $\mathrm{^{39}K}$. The effective detuning $\delta_{\mathrm{eff}}$ of $\mathrm{\mathcal{L}_{\mathrm{sl}}}$ together with the applied magnetic field B dictates which velocity class is adressed (see relation \ref{relation}). \textbf{(b)} Sideband spectrum of the slowing Laser (black) to adress the transition frequencies for the different hyperfine levels in high magnetic fields.\textbf{(c)} Spectrum of the frequency broadened repumping laser realized with a current modulated DFB Diode. Sinusoidal modulation with $f_{\mathrm{mod}}=\mathrm{12\,MHz}$ modulated to a width of approximately $\mathrm{900\,MHz}$.\label{slowingscheme}}
\end{figure*}
\end{center}

Type-\Romannum{2}  Zeeman slowing has been presented in  \cite{petzold_zeeman_2018} focussing on a discussion of possible Zeeman slowing of diatomic molecules and a brief presentation of a proof-of-principle experiment working on the  $\mathrm{D_{1}}$-line of atomic $^{39}$K. Here, we will briefly explain the scheme making use of the example of atomic $^{39}$K.

Figure \ref{slowingscheme} shows the level structure of the D$_{1}$-line in an external magnetic field. For magnetic fields leading to Zeeman shifts larger than the hyperfine splitting the system enters the Paschen-Back regime and the level structure is largely simplified. The ground and excited states each split into two manifolds with $m_{J}=\pm 1/2$. Neglecting hyperfine structure this can be seen as an effective 4-level system (see Fig. \ref{slowingscheme} \textbf{(a)}). One of the ground state manifolds is then coupled to an excited state  manifold $\left| m_J=\pm 1/2 \right\rangle \rightarrow \left| m_{J'}=\mp 1/2 \right\rangle $  by a laser $\mathcal{L}_{\mathrm{sl}}$ with $\sigma^{\mp}$ polarization. The corresponding transition frequency can be tuned by a magnetic field and can therefore be used to compensate a changing Doppler shift during the deceleration process in either an increasing ($\sigma^{-}$) or an decreasing ($\sigma^{+}$) field Zeeman slower configuration over the relation:

\begin{align}
\delta_{\mathrm{eff}}+(g_{J}\cdot m_{J}-g_{J'}\cdot m_{J'})\cdot \mu_{B}\cdot B+k\cdot v&=0 \\
\delta_{\mathrm{eff}}+\mu_{\mathrm{eff}}\cdot B + k\cdot v &= 0 \label{relation}
\end{align} 

Here $g_{J}$ and $g_{J'}$ are the $g_{J}$-factors in the Paschen-Back regime of the ground state $4^{2}S_{1/2}$ and the excited state $4^{2}P_{1/2}$ respectively. As the transition between these manifolds is not closed the atom will quickly fall into the ground state manifold, which is not adressed by $\mathcal{L}_{\mathrm{sl}}$. 
Additionally this non coupled manifold exhibits a different Zeeman shift and therefore cannot be pumped back by the application of a simple sideband. Still a frequency broadened laser $\mathcal{L}_{\mathrm{rep}}$ (see Fig. \ref{slowingscheme}\textbf{(a)} and \textbf{(c)}) can be applied covering all relevant velocity classes for the applied magnetic fields to pump the atoms from the dark manifold back to the one addressed by $\mathcal{L}_{\mathrm{sl}}$.

At the beginning of the slowing process all atoms are quickly pumped into the ground state manifold adressed by $\mathcal{L}_{\mathrm{sl}}$. They stay in this state until the magnetic field reaches a magnitude that shifts the level, at a given velocity of the atom, into resonance with $\mathcal{L}_{\mathrm{sl}}$.
The slowing force itself is then exerted by subsequent scattering of photons by $\mathcal{L}_{\mathrm{sl}}$ and $\mathcal{L}_{\mathrm{rep}}$ with equal amount. Note however, that the resonance condition for the whole system is still governed by $\mathcal{L}_{\mathrm{sl}}$ so that the system acts similar to a traditional type-\Romannum{1} Zeeman slower. At the end of the slowing process the atoms are again pumped by $\mathcal{L}_{\mathrm{rep}}$ into the manifold formerly adressed by $\mathcal{L}_{\mathrm{sl}}$.

To take the finite hyperfine structure of $^{39}$K in the ground and the excited state into account, $\mathcal{L}_{\mathrm{sl}}$ needs 4 sidebands to couple each hyperfine state $m_{I}$ to the corresponding hyperfine level in the excited state (see Fig. \ref{slowingscheme} \textbf{(b)}). Note that for every specific hyperfine state $m_{I}$ on its own, there are 4 velocity classes resonant with $\mathcal{L}_{\mathrm{sl}}$ due to its 4 sideband structure. However, mixing between different $m_I$ states is still significant at the magnetic fields we work at, so that the 4 sidebands do not result in four distinct slowing peaks in our velocity distribution. 

\begin{center}
\begin{figure*}
\onecolumngrid
	\includegraphics[width=0.9\textwidth]{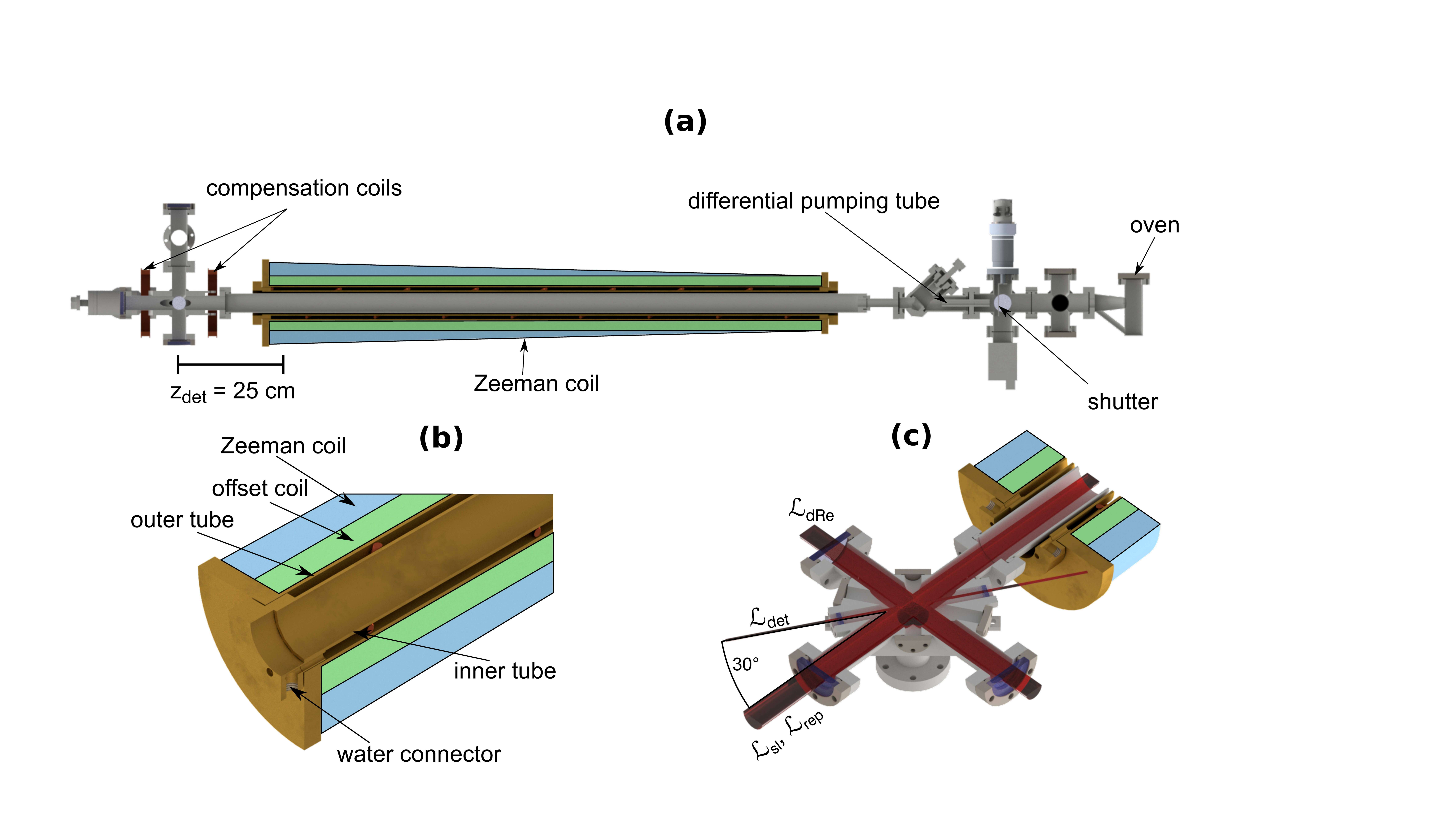}%

	\caption{\textbf{(a)} Half section view of the vacuum chamber showing the overall geometry of the apparatus. \textbf{(b)} Quarter section view of the magnetic field coil. Between the inner and the outer tube of the brassholder cooling water is flushed, which holds the coil at a temperature of $\mathrm{T<70^{\circ}C}$. \textbf{(c)} Detection chamber showing the slowing and detection lasers. For the fluorescence measurement the lens and curved mirror are additionally installed in this chamber (not shown here).\label{apparatus}}
\end{figure*}
\end{center}

%
%
%

\section{Experimental apparatus}
In our experiments, we realize a type-\Romannum{2} Zeeman slower for atomic potassium based on the apparatus sketched in  Fig. \ref{apparatus}. The atomic beam is created in an oven which is heated to $\mathrm{190^{\circ}C}$, Fig. \ref{apparatus}\textbf{(a)}. The atoms leave the oven through a $\mathrm{3\,mm}$ skimmer 
and enter the $130\,$cm long Zeeman slowing region. Here, the atoms interact with the counter propagating slowing light generated by a self-built diode laser system. The  spatially varying Zeeman field is produced by  two  carefully designed coils.  Finally, the slowed atoms reach the detection region $\mathrm{25\,cm}$ behind the end of the slowing region. Here, the longitudinal atomic velocity after the slowing process is analyzed via  Doppler spectroscopy.

\subsection{Magnetic field}
 In our experiment we decided for an increasing field Zeeman slower for which the ideal magnetic field is given by 

\begin{align}
B(z)&=B_{0}+\Delta B\cdot (1-\sqrt{1-z/L_{0}}) \text{ for } 0<z<z_{\mathrm{fin}}
\end{align}

Here $B_{0}$ denotes the offset field, which in the case for type-\Romannum{2} Zeeman slowing is large enough to bring the particles in the Paschen-Back regime.
$L_{0}$ is the length of the slowing field needed to slow down the atoms from the capture velocity  $v_{\mathrm{cap}}$ to a standstill given by 

\begin{align}
L_{0}=m\frac{v_{\mathrm{cap}}^{2}}{\eta \hbar k \Gamma}
\end{align}

where m is the mass of the particle, $\Gamma$ is the spontaneous decay rate of the excited state and $\eta=\frac{a}{a_{\mathrm{max}}}$ is the so called design parameter, which dictates with how much of the maximum attainable deceleration $a_{\mathrm{max}}=\frac{\hbar k \Gamma}{2 m}$ the slower is operated.
The ideal field for a deceleration to a specific final velocity $v_{\mathrm{fin}}$ stops at

\begin{align}
z_{\mathrm{fin}}=m\frac{v_{\mathrm{cap}}^{2}-v_{\mathrm{fin}}^{2}}{\eta \hbar k \Gamma}
\end{align} 

Together with the effective detuning of the slowing laser sidebands $\delta_{\mathrm{eff}}$ it dictates the capture velocity of the Zeeman slower due to the relation $\mu_{\mathrm{eff}}\cdot B_{0}+k\cdot v_{\mathrm{cap}}=-\delta_{\mathrm{eff}}$. The final velocity is given by the magnetic field at the end of the slowing region $B(z_{\mathrm{fin}})$, through $\mu_{\mathrm{eff}}\cdot B(z_{\mathrm{fin}})+k\cdot v_{\mathrm{fin}}=-\delta_{\mathrm{eff}}$. 

 In our experiments the solenoid producing the  Zeeman field consists of two independent coils. The inner offset coil in Fig. \ref{apparatus}\textbf{(b)} produces a homogeneous magnetic field over the entire slower length serving as an offset field to access the Paschen-Back regime. The second coil produces a typical increasing field Zeeman slower field. The sum of the two fields gives a good approximation of the ideal magnetic field and the separation of coils gives us the capability to tune the capture velocity $v_{\mathrm{cap}}$ and the final velocity $v_{\mathrm{fin}}$ independently.
 
During the experiment the solenoid dissipates a power of approximately $\mathrm{850\,W}$ and therefore needs to be watercooled over its entire length. Thus it consists of two coaxial brass tubes, which are stacked into each other and are sealed to each other on both sides, see also Fig. \ref{apparatus}\textbf{(b)}). Through a water connector we flush cooling water through the whole brassholder.  During the experiment the coil reaches a temperature of $\mathrm{60^{\circ}C}$ on the surface and $\mathrm{70^{\circ}C}$ on the estimated hottest point in between the windings, where a temperature sensor was placed during the construction. The wire is rated up to a temperature of $\mathrm{200^{\circ}C}$, so even higher fields are possible with this design. 

Two compensation coils installed around the detection region are used to pull the magnetic field to $\mathrm{B=0\,G}$ and the magnetic field gradient to $\mathrm{\frac{dB}{dz}=0\,G \,cm^{-1}}$.


%

\subsection{Laser systems}
The necessary laser systems are realized by means of diode laser systems. An external cavity diode laser system (ECDL) is used as $\mathcal{L}_{\mathrm{sl}}$ to couple the $\mathrm{4^{2}S_{1/2}\left|m_{J}=1/2,m_{I}=-3/2,...,+3/2 \right\rangle}$ ground state sublevels to the $\mathrm{4^{2}P_{1/2}\left|m_{J}=-1/2,m_{I}=-3/2,...,+3/2 \right\rangle}$ excited state sublevels with light polarized to drive $\sigma^{-}$-transitions. $\mathcal{L}_{\mathrm{sl}}$ is locked via modulation-transfer-spectroscopy $1.618\,\mathrm{GHz}$ to the red of the $\mathrm{D_{1}}$-line-crossover in a $\mathrm{^{39}K}$ spectroscopy cell. The light is amplified by a tapered amplifier and afterwards passes through a double-pass acousto-optic modulator system (AOM) to create the 4 sidebands needed to address the 4 hyperfine levels $\mathrm{m}_{I}=-3/2,...,+3/2$. The AOM is driven at a frequency of $118.6\, \mathrm{MHz}$ and we combine the $\mathrm{m=+2,+1,0,-1}$ orders with a 50:50 beamsplitter, as we need all frequencies with the same polarization for the later combination with $\mathrm{\mathcal{L}_{rep}}$. $\mathcal{L}_{\mathrm{sl}}$  is coupled into a polarization-maintaining single-mode fiber to clean the beamprofile.

Light for $\mathcal{L}_{\mathrm{rep}}$ is produced by a tapered amplifier, seeded with a current modulated DFB diode to frequency broaden the light.
For the shown measurements the diode is modulated sinusoidally with a modulation frequency  $f_{mod}=12\, \mathrm{MHz}$ to a width of $\Delta f \approx 1.6\,\mathrm{GHz}$. In Fig. \ref{slowingscheme}\textbf{(c)} the frequency spectrum of $\mathcal{L}_{\mathrm{rep}}$ is shown
recorded by a cavity with a free spectral range of $1\, \mathrm{GHz}$ when modulated to a width of $\Delta f \approx 900 \, \mathrm{MHz}$. 
The spatial transverse beam profile is cleaned by coupling through a polarization-maintaining single-mode fiber.

After combining $\mathrm{\mathcal{L}_{sl}}$ and $\mathrm{\mathcal{L}_{rep}}$ with a polarizing beamsplitter we use a quarter-wave plate to circularly polarize the lasers. We end up with a power of $\mathrm{P_{rep}=400\,mW}$ in total and $\mathrm{P_{sl}\approx 20\,mW}$ in each order. The beams have a Gaussian beamwaist of $w_{0}=1\,\mathrm{cm}$ at the vacuum viewport and are slightly focussed by an adjustable telescope towards the oven region.


\subsection{Velocity measurement} The resulting atomic velocity distribution is analyzed via Doppler spectroscopy. A detection laser $\mathrm{\mathcal{L}_{det}}$ intersects the atomic beam   at an angle of $30^{\circ}$ (see Fig. \ref{apparatus} \textbf{(c)}). The detection takes place on the D$_{2}$-Line of $^{39}$K from the $4^{2}S_{1/2},F=2$ state to the $4^{2}P_{3/2}$ state, where we do not resolve the hyperfine structure of the excited state resulting in a velocity resolution of $26 \, \mathrm{m \, s^{-1}}$. Additionally, a repumping laser $\mathrm{\mathcal{L}_{dRe}}$ intersects the atomic beam in the detection region at an angle of $90^{\circ}$ pumping all atoms from the $4^{2}S_{1/2},F=1$ to the $4^{2}S_{1/2},F=2$ detection state and is therefore not Doppler sensitive.$\mathrm{\mathcal{L}_{det}}$ is offset locked to a master laser stabilized via frequency-modulation spectroscopy on a potassium spectroscopy cell. Typically we scan $\mathrm{\mathcal{L}_{det}}$ linearly to beat frequencies of up to $\mathrm{1 \, GHz}$, which corresponds to a velocity of $\mathrm{700\, m\,s^{-1}}$ taking into account the locking point of the master laser and the geometry of the detection. Most of our data we record in absorption. Since $\mathrm{\mathcal{L}_{det}}$ has a relative intensity noise of $10^{-2}$ and the absorption signal is in the $10^{-4}$ range we use a differential absorption scheme for most of our measurements. We later additionally installed a fluorescence detection system with inside vacuum optics for the white-light slowing measurement.
We determine the velocity axis by first measuring the absorption signal at an angle of $\mathrm{90^{\circ}}$ to the atomic beam, the peak position gives us the beat frequency which corresponds to $\mathrm{0\,m\,s^{-1}}$. The scaling of the axis is then given by the measured beat frequency the angle of  $\mathrm{\mathcal{L}_{det}}$ and the corresponding Doppler shift.

\section{Results}

\begin{figure}
	\includegraphics[width=0.43\textwidth]{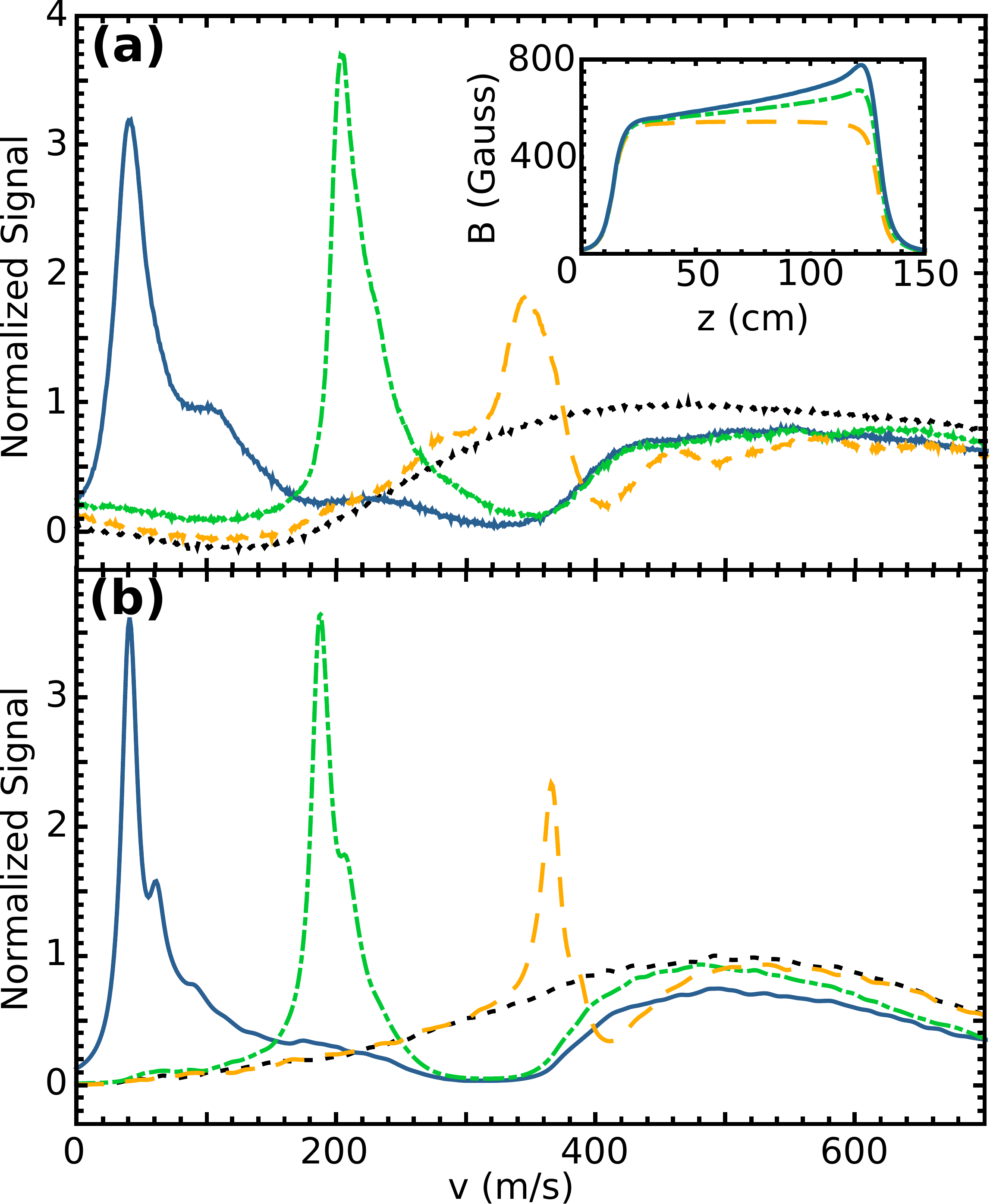}%

	\caption{\textbf{(a)} Differential absorption signals for type-\Romannum{2} Zeeman slowing for several final velocities. The inset shows the corresponding configurations of the applied magnetic field. \textbf{(b)} Corresponding three dimensional Monte Carlo simulations. The dotted black curve shows the initial velocity distribution. When only the offset field is applied (dashed orange curve) the slowing results in a final peak velocity of $v_{\mathrm{p}}\approx 350 \, \mathrm{m \, s^{-1}}$. By adding a Zeeman Field (dot-dashed green and solid blue curves) the slowing laser is kept in resonance with the slowed $^{39}$K resulting in a shift of $v_{\mathrm{p}}$ to lower velocity classes. The simulations are in good agreement with the measured results but typically predict slightly narrower velocity peaks.   \label{difffinal}}
\end{figure}

Figure \ref{difffinal} shows the measured velocity distributions after Zeeman slowing for different magnetic field profiles. First, we measured the initial velocity distribution (dotted black line) with no slowing lasers applied. We use the height of the initial distribution peaking at approximately $514\mathrm{\,m\,s^{-1}}$ to normalize our signals. We then switch on the slowing lasers to allow for radiative slowing. The dashed orange graph was recorded with only the offset field on (see the inset in Fig. \ref{difffinal}\textbf{(a)} for the corresponding magnetic fields). Here the atoms are slowed down from $400\mathrm{\,m\,s^{-1}}$ to approximately $350\mathrm{\,m\,s^{-1}}$ at which point the atoms are slowed down so far that $\mathcal{L}_{\mathrm{sl}}$ is not resonant anymore. The subsequent graphs show, that by adding a Zeeman field $\Delta B$ this peak can be shifted downwards to lower velocities. The solid blue graph shows the distribution when slowed to a final peak velocity of $v\mathrm{_{p}=35\,m\,s^{-1}}$. The simulated height and position of the final velocity peaks in Fig. {\ref{difffinal}}\textbf{(b)} are in good agreement with the measured results. We feed the measured initial velocity distribution into the simulation and take transverse heating effects due to the spontaneous emission and the Gaussian intensity profile of the slowing lasers into account. Small differences for $v_{\mathrm{p}}$ are most probably due to uncertainties in the applied magnetic fields.

\begin{figure}
	\includegraphics[width=0.43\textwidth]{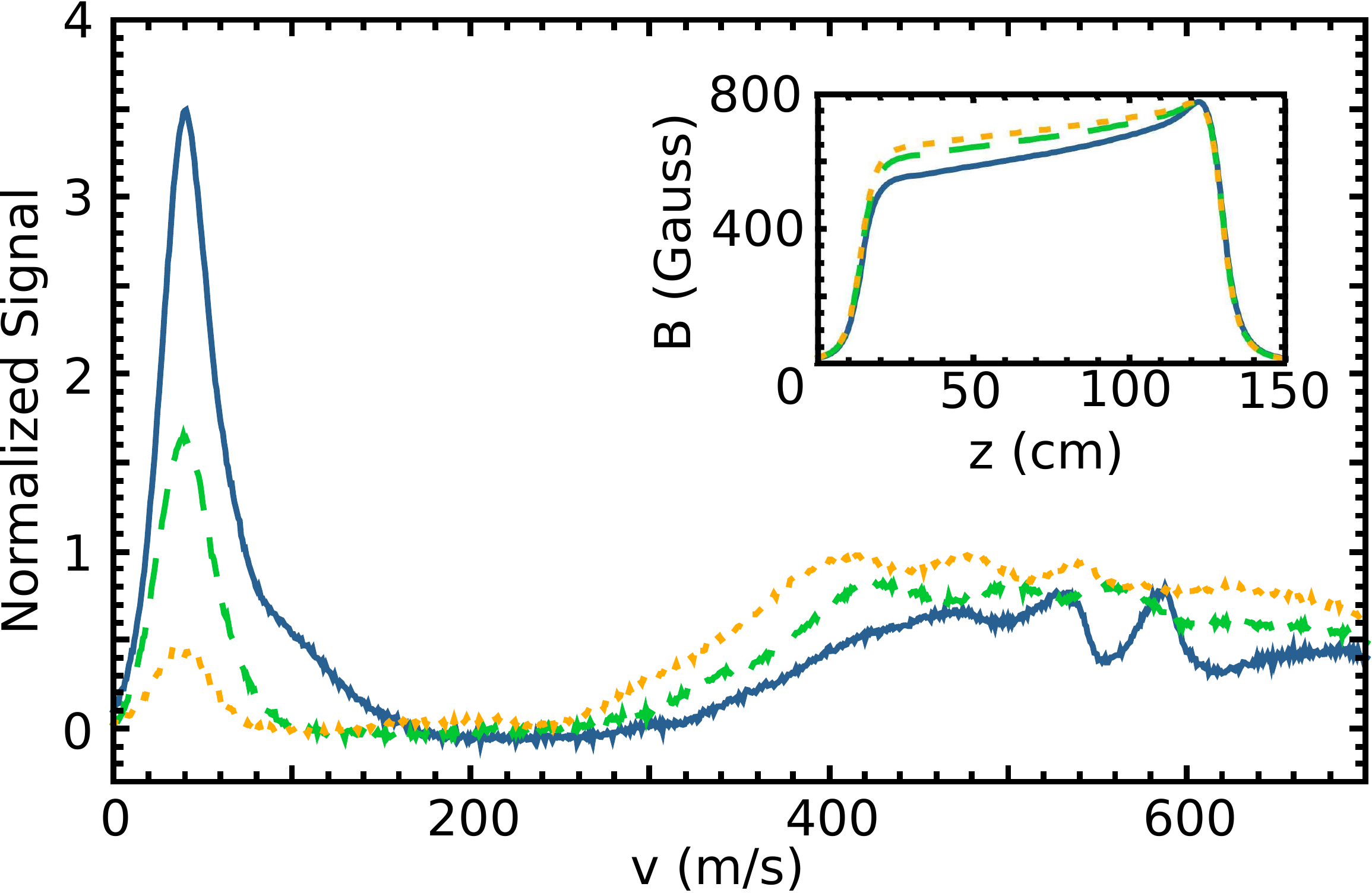}%

	\caption{Type ll Zeeman slowing with different capture velocities. By adressing higher velocity classes while keeping the final velocity constant the slowing performance can be optimized.For the dotted orange curve $\Delta B\approx 147 \, \mathrm{G}$ corresponds to a velocity change of $\Delta v \approx 211 \, \mathrm{m \, s^{-1}}$. The solid blue curve shows the optimized result with  $\Delta B \approx 234 \, \mathrm{G}$ corresponding to $\Delta v \approx 335 \, \mathrm{m \, s^{-1}}$ such that significantly more velocity classes are being slowed and compressed into the final velocity peak. \label{diffcap}}
\end{figure}

In a next step, we optimize the performance of the Zeeman slower by tuning the capture velocity. In Fig. \ref{diffcap} we keep the maximum magnetic field constant and alter the magnetic field magnitude at the entrance of the slower.  For a small capture velocity only a few atoms are swept from higher velocities down to $v\mathrm{_{p}=40\,m\,s^{-1}}$. If we decrease the magnitude at the entrance of the slower further, we can increase the flux at low velocities by capturing more atoms.


\begin{figure}

	\includegraphics[width=0.43\textwidth]{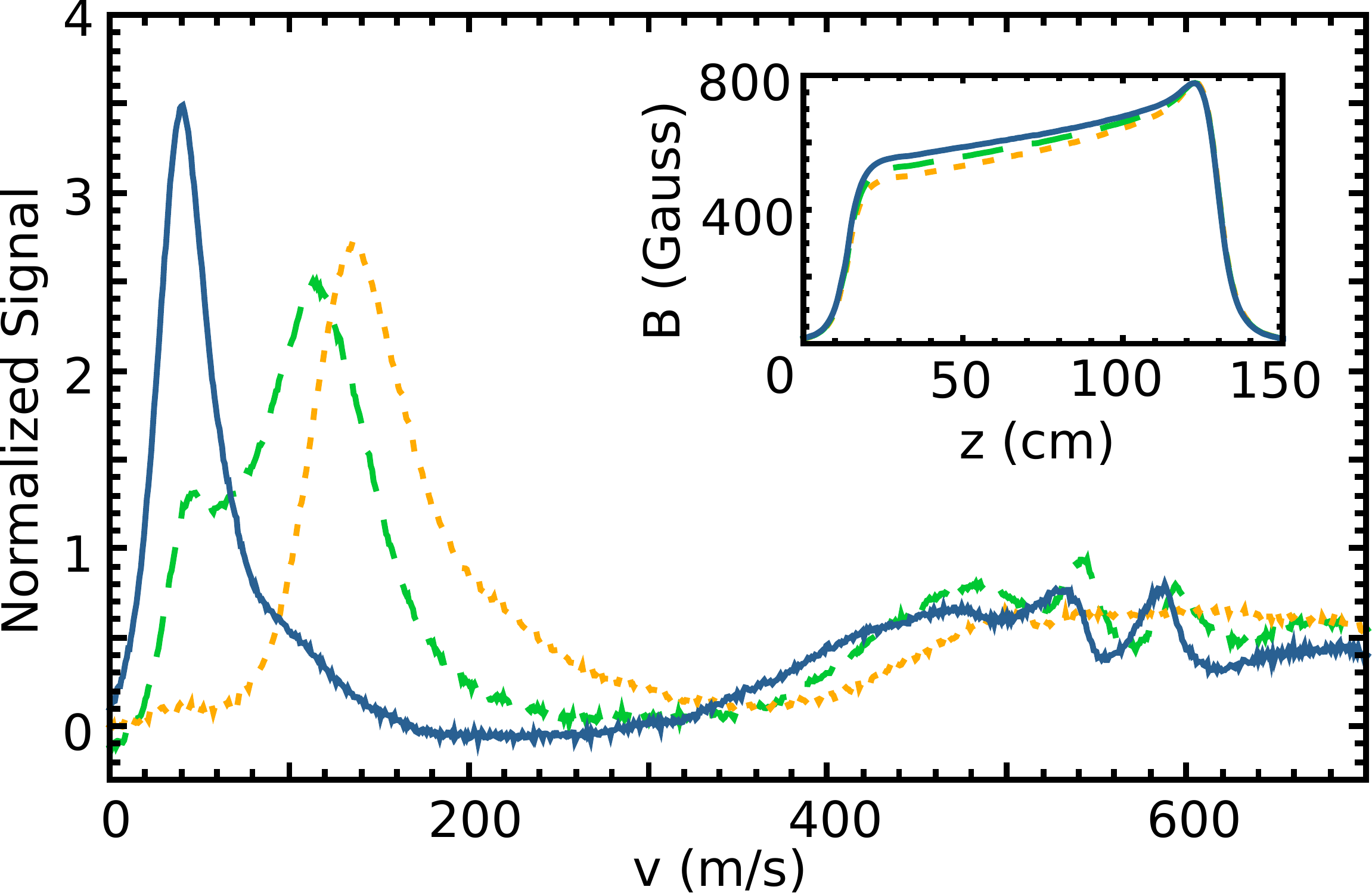}%

	\caption{In solid blue the optimized signal is shown alongside with the corresponding magnetic field for our Type ll Zeeman slower. The dashed green and dotted orange lines show the results, when the capture velocity is increased even further with capture velocities too high to be slowed over the finite slowing path. The atoms fall out of resonance with $\mathcal{L}_{\mathrm{sl}}$ before reaching their final velocity. For the dashed green curve, atoms travelling at the centerline of the slowing lasers are still slowed to the desired final velocity, atoms travelling off center fall out of resonance with the slowing beam. This is due to the Gaussian beam profile of the slowing laser having a significantly lower intensity in the wings 
	For the dotted orange curve the magnetic field gradient is so steep that all atoms fall out of resonance with the slowing laser at higher velocity classes.  \label{toomuch}}
\end{figure}

Our slowing performance is limited by the deceleration $a=\eta a_{\mathrm{max}}$  we achieve with the limited repump power in our experiment. This can be seen
when increasing the capture velocity even further. The system then reaches a point where the atoms cannot follow the changing resonance condition dictated by the magnetic field gradient anymore (see Fig. \ref{toomuch}). Then they fall out of resonance with $\mathcal{L}_{\mathrm{sl}}$ before reaching $v_{\mathrm{fin}}$. The dashed green measurement shows a double peak structure,
resulting from the fact that the achieved deceleration that $\mathrm{\mathcal{L}_{rep}}$ can provide is different for particles on-axis versus off-axis due to its Gaussian beam profile.
On axis, where the intensity is highest, the slowing still takes place and the atoms are slowed down to $v\mathrm{_{fin}=40\,m\,s^{-1}}$ but in the wings of the profile the intensity is not sufficient and they fall out of resonance with $\mathcal{L}_{\mathrm{sl}}$ at higher velocities. We checked this interpretation by placing an iris in front of the entrance vacuum view port of the slowing and repumping laser. By cutting the wings of $\mathcal{L}_{\mathrm{sl}}$ and $\mathcal{L}_{\mathrm{rep}}$ the peak at higher velocities disappears and only the peak at $v\mathrm{_{fin}=40\,m\,s^{-1}}$ was left. We see the same behaviour in our Monte-Carlo simulations (not shown here).

In our measurements, we find the type-\Romannum{2} Zeeman slower system to work reliably with results easily reproduced on a daily basis. When optimizing the system we found some parameters to be of importance, especially the beam overlap between $\mathcal{L}_{\mathrm{rep}}$ and $\mathcal{L}_{\mathrm{sl}}$ is crucial. Slightly focussing the lasers throughout the slowing region on the oven nozzle is benefitial due to a small transverse confinement force, but the gain is quite moderate.
While the power of $\mathcal{L}_{\mathrm{rep}}$ is the limiting factor in our experiment the exact modulation frequency seems to be quite unimportant. We get our best slowing performance for sinusoidal modulation with $\mathrm{f_{mod}=12\,MHz}$. We additionally tested modulation frequencies ranging from $\mathrm{f_{mod}=6-24 \, MHz}$ as well as other modulation schemes like sawtooth and triangular modulation, which altered the frequency spectrum significantly but changed the slowing result by only 20 to $30 \, \%$.
Overall we attained a flux of $\Phi_{\mathrm{type \Romannum{2}}}=3.3 \cdot 10^{9}\,\mathrm{cm}^{-2}\mathrm{s}^{-1}$ under $35 \, \mathrm{m \, s^{-1}}$ for an oven temperature of $T_{\mathrm{oven}}=190 ^{\circ}C$. Due to the available repump power limiting our slowing performance, we reach a fraction of $\eta=\frac{a}{a_{\mathrm{max}}}=0.38$ of the maximum attainable deceleration $a_{\mathrm{max}}$.



\section{Comparison to conventional radiative beam slowing methods}
\label{typeone}
\subsection{Type-\Romannum{1} Zeeman slower}

First, we compare to the slowing result of a traditional type-\Romannum{1} Zeeman slower working on the D$_{2}$-line of $^{39}$K. These are a standard beam slowing method used to load magneto-optical traps for alkali or alkaline earth atoms and can therefore be seen as a benchmark for our type-\Romannum{2} Zeeman slower. The maximum achievable radiation pressure on the D$_{2}$-line $(\lambda=766.7 \, \mathrm{nm}, \, \tau = 26.37 \, \mathrm{ns})$ is approximately the same as on the D$_{1}$-line $(\lambda=770.1 \, \mathrm{nm} , \, \tau = 26.72 \, \mathrm{ns})$ and the comparison gives a good hint if the velocity selectivity and the distribution of the force along the slowing path for our type-\Romannum{2} Zeeman slower is as favorable as in the type-\Romannum{1} case. To our best knowledge this is the first time a Zeeman slower is implemented on the D$_{2}$-line of $^{39}$K. 

\begin{figure}
	\includegraphics[width=0.43\textwidth]{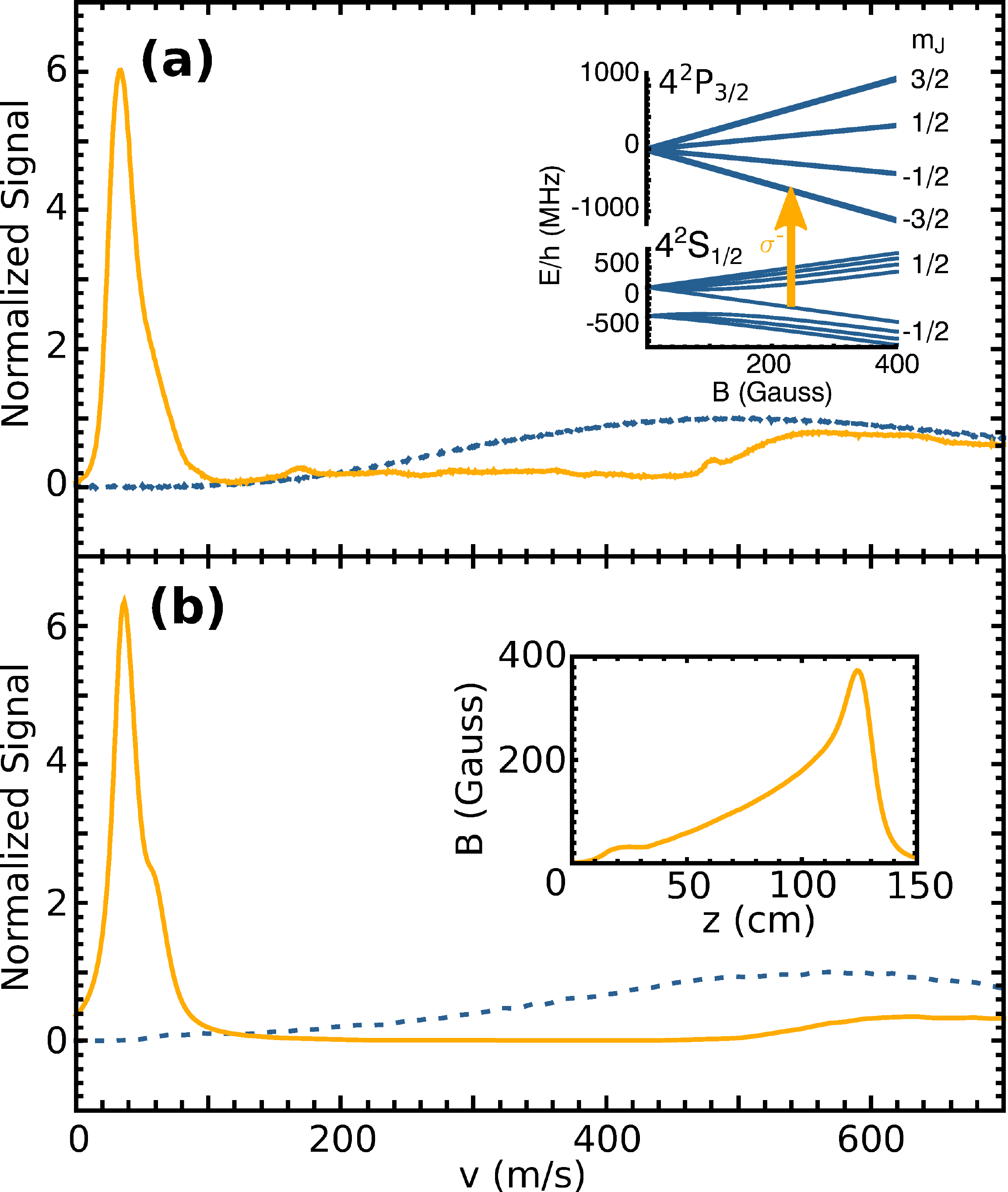}%

	\caption{\textbf{(a)} Differential absorption signal of type-\Romannum{1} Zeeman Slowing in a $\mathrm{\sigma^{-}}$-configuration (see inset). The dotted blue curve shows the initial velocity distribution, whereas the solid orange curve shows the slowed distribution.  \textbf{(b)} Corresponding Monte Carlo simulation. The inset shows the applied magnetic field. We do not use an offset field for the type-\Romannum{1} slower. \label{dtwoslower}}
\end{figure}

To realize a type-\Romannum{1} Zeeman slower in our setup, we lock $\mathrm{\mathcal{L}_{sl}}$  $\mathrm{809\,MHz}$ red of the D$_{2}$-line crossover. The light is polarized to drive $\sigma^{-}$-transitions so that the slowing takes place on the closed $\mathrm{4^{2}S_{1/2},\left|F=2,m_{F}=-2\right\rangle \rightarrow 4^{2}P_{3/2},\left|F=3,m_{F}=-3\right\rangle}$ transition as shown in the inset of Fig. \ref{dtwoslower}. 
No sidebands for $\mathcal{L}_{\mathrm{sl}}$ and no repumper is needed in type-\Romannum{1} Zeeman slowing. 
We get the best slowing result for our type-\Romannum{1} Zeeman slower when no offset field $\mathrm{B_{offset}=0 \, G}$ is applied, as shown in the inset of Fig. \ref{dtwoslower}.
The magnetic field is generated by the shaped magnetic field coil resulting in an increasing field Zeeman slower starting with $B=0 \, \mathrm{G}$ and ending at about $B=375 \, \mathrm{G}$ at the end of the slowing region.

Figure \ref{dtwoslower} \textbf{(a)} shows the slowing result we achieved in our experiment. The capture velocity lies at $v_{\mathrm{cap}} \approx 500 \, \mathrm{m \, s^{-1}}$ and the final peak velocity at $v_{p}\approx 35 \, \mathrm{m \, s^{-1}}$. Between these two velocities there is nearly no atom signal showing the efficient deceleration over the whole velocity range. In Fig. \ref{dtwoslower} \textbf{(b)} the corresponding Monte Carlo simulation is shown, again exhibiting a good agreement with the experimental results.
We find that the type-\Romannum{1}  slower is more dependent on exact beam alignment and focussing conditions of the slowing beams than in the type-\Romannum{2} case. A focus near the nozzle of the oven leads to high intensities where the resulting power broadening of the transition is large enough to cover the whole Doppler range of the longitudinal velocity distribution  \cite{molenaar_diagnostic_1997}. As a result all atoms are pumped into the stretched slowing state.  In contrast to this, in type-\Romannum{2} Zeeman slowing all ground state sublevels are coupled, so that there is no need for optical pumping into a desired state.
 Although it has been noted elsewhere \cite{hopkins_versatile_2016, wille_preparation_2009} that the small hyperfine splitting in the excited state could lead to loss from the slowing process,  we do not find this to be a problem.


We measure a flux of atoms  with velocities slower than 35$\mathrm{\,m\,s^{-1}}$ of $5.5 \cdot 10^{9} \, \mathrm{cm^{-2}s^{-1}}$.  This is a factor of 1.66 larger than the flux measured in the type-\Romannum{2} Zeeman slower scheme. The better performance arises from the higher capture velocity of about $500 \mathrm{\,m\,s^{-1}}$ instead of $350\mathrm{\,m\,s^{-1}}$ in the type-\Romannum{2} case, the latter being limited by more demanding laser power requirements.
 Furthermore, the type-\Romannum{1} slower effectively starts slowing in the oven region, where for the case of the type-\Romannum{2} slower the actual slowing starts not until the field reaches $B_{0}$ inside the solenoid. 
The width of the  resulting velocity distribution  for the type-\Romannum{1} and type-\Romannum{2} Zeeman slower are comparable and  limited by the resolution of our detection system in both cases. Indicating that the cutoff of the slowing force in the rapidly decreasing magnetic field is as sharp for the type-\Romannum{2} slower as in the type-\Romannum{1} case and therefore the slowing and compression characteristics are similar.

\subsection{White-light slowing}

In a next step, we compare type-\Romannum{2} Zeeman slowing to white-light slowing, which is an established method for radiative slowing of a molecular beam \cite{barry_laser_2012, hemmerling_laser_2016}. 

\begin{figure}
	\includegraphics[width=0.43\textwidth]{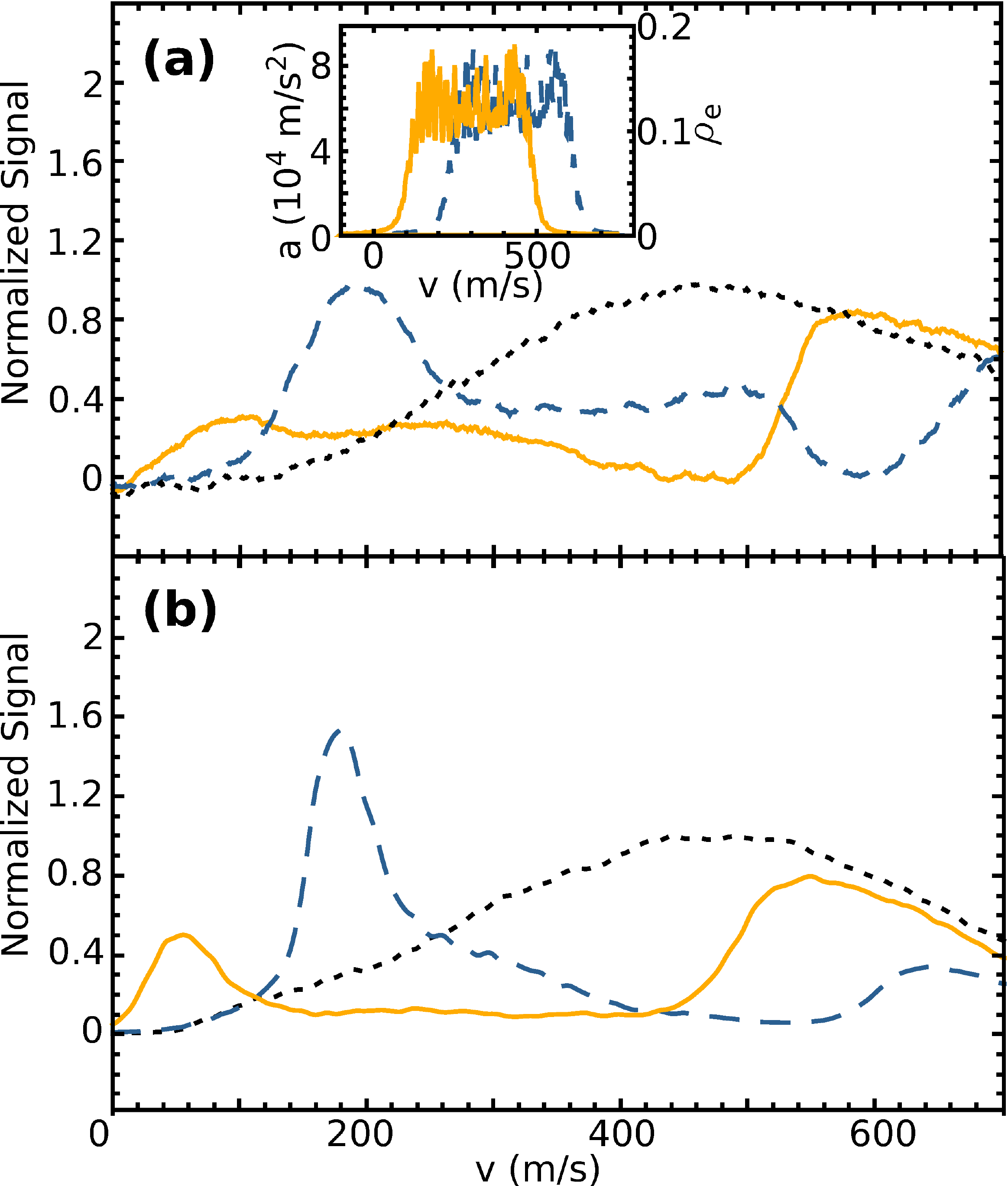}%

	\caption{\textbf{(a)} Achieved white-light slowing signals for different $v_{\mathrm{fin}}$. The dotted black line shows the initial velocity distribution. The dashed blue and solid orange curves show the slowed velocity distributions and the inset shows the corresponding force profiles calculated by multilevel rate equations spanning over a velocity range of $\Delta v \approx 400 \, \mathrm{m \, s^{-1}}$. The density at low velocity classes increases when the slowing force is applied, but the efficiency is rapidly decreasing when slowing to low velocities. \textbf{(b)} The corresponding Monte Carlo simulations show a similar behaviour but in the simulations there are nearly no atoms lost at higher velocity classes, as can be seen for the solid orange curve in \textbf{(b)}. This is most probably due to a lower deceleration force in the experiment in comparison to the predicted force. \label{WLFluo}}
\end{figure}

In our setup, frequency broadened laser light is provided by $\mathcal{L}_{\mathrm{rep}}$, this time detuned to the red side of the potassium resonance. We modulate with a sinusoidal frequency of $f_{\mathrm{mod}}=12\,\mathrm{MHz}$ to an approximate width of $\Delta \omega\approx 900\,\mathrm{MHz}$ corresponding to a velocity range of $\Delta v \approx 400 \, \mathrm{m \, s^{-1}}$.

 The laser is linearly polarized and
 we find the best slowing result when a magnetic field of $\mathrm{B_{0}}=11\,\mathrm{G}$ is applied throughout the slowing region. We suppose this to be due to enhanced destabilization of coherent dark states on the D$_{1}$-line in a magnetic field \cite{berkeland_destabilization_2002}.   The final velocity distribution of the atoms after slowing is measured by detecting the fluorescence induced by $\mathcal{L}_{\mathrm{det}}$. Although the resulting signals are usually noisier and need more time to average than the differential absorption it was used here because of two reasons. First, the white-light slowing lasers are close to resonance in the detection region and pump the atoms into the $4^{2}S_{1/2},\left|F=1\right \rangle$ state. We therefore would need a high power repumper $\mathcal{L}_{\mathrm{dRe}}$ to pump the atoms back into the $4^{2}S_{1/2},\left|F=2\right \rangle$ detection state over the whole region where $\mathcal{L}_{\mathrm{det}}$ crosses the atomic beam.
Because the flourescence optics only collects light from a small volume in the middle of the chamber, the power of $\mathcal{L}_{\mathrm{dRe}}\approx 20 \, \mathrm{mW}$ is sufficient to pump the atoms back here.
 The second reason is background potassium gas, which we see more in the absorption pictures than in the fluorescence pictures, as the absorption laser crosses the whole chamber. This effect is of no importance for the Zeeman slowing data as the slowing peaks are much higher, but for the small signals at low velocities in white-light slowing it heavily disturbs the slowing result.

 Figure \ref{WLFluo}\textbf{(a)} shows the measured velocity distributions.
 For the different measurements we kept the modulation of the slowing laser constant and altered the offset detuning resulting in different final velocities.
 
The initial velocity distribution is shown as the dotted black curve.
 For the dashed blue graph we see a deceleration from approximately $600\, \mathrm{m\,s^{-1}}$ down to $200\, \mathrm{m \, s^{-1}} $. The slowing laser cuts a deep hole into the velocity distribution and increases the density at $200\, \mathrm{m\, s^{-1}}$ significantly. If we detune the slowing laser to be resonant with lower velocity classes, the capture velocity and the final velocity shift accordingly (solid orange curve). The density at lower velocities still increases in comparison to the unslowed distribution but the efficiency is rapidly decreasing. This behaviour is seen in many publications regarding white-light slowing for atoms \cite{chae_laser_2015} and molecules \cite{barry_laser_2012, truppe_intense_2017, hemmerling_laser_2016}. As it was already argued in Fig. \ref{SlowingMethods} for white-light slowing all atoms reach their final velocity at a different point in space. Therefore the time the atoms spend in the slowing region is not minimized as in type-\Romannum{1} or type-\Romannum{2} Zeeman slowing and the transverse spread leads to high losses until the slowed velocity class finally reaches the detection region. The second effect is that the slowing force has no sharp cutoff at low velocities,due to the finite hyperfine structure of the excited state, the natural linewidth of the slowing transition, which is additionally power broadened, and the fact, that the  frequency spectrum of the slowing laser as seen in Fig. \ref{slowingscheme}\textbf{(c)} has no sharp cutoff at high frequencies. Atoms  which have already reached the desired end velocity will still be decelerated while flying to the detection region, leading to spreading of the slowed velocity peak. This effect gets stronger the slower the atoms are. For very low velocities this can even lead to deceleration under $0\, \mathrm{m\,s^{-1}}$ where the atoms obviously never reach the detection region ( see also Fig. \ref{SlowingMethods}).
 In Fig. \ref{WLFluo}\textbf{(b)} the corresponding Monte Carlo simulations are shown and the inset of Fig. \ref{WLFluo}\textbf{(a)} shows the calculated deceleration force used for the simulation. The force is calculated by multilevel rate equations, taking into account the frequency spectrum of the slowing laser and the hyperfine structure of the D$_{1}$-line  resulting in a system with 8 ground state and 8 excited state sublevels. The slowing laser in the simulation is polarized to drive $\sigma^{+}$ and $\sigma^{-}$ transitions with the same amplitude. This is realized in the experiment through the linearly polarized laser and the longitudinal magnetic offset field.
  
 The simulations show good agreement with the measured distributions concerning the capture and the final velocity. Also the peak at very slow velocities is small compared to the Zeeman slowing cases as the efficiency of the slowing to low velocity classes rapidly decreases. Still the results from the simulation suggest a slightly better slowing result as was measured. In the simulation nearly no atoms are lost at higher velocities (see the solid orange curve in Fig. \ref{WLFluo}\textbf{(b)}). Presumably due to a smaller force in the experiment than the rate equation model suggests. In fact we measure the best slowing result with a small magnetic offset field applied supposedly due to remixing of coherent dark states. This effect is not taken into account in the rate equation model and could explain the lower force and therefore the loss at higher velocities in the experiment. A more realistic force profile could be obtained by solving the optical bloch equations for the whole system.
  
 In the white-light slowing case the laser does not have to be as broad in frequency ($\Delta f \approx 900 \, \mathrm{MHz}$) as the repump laser for the type-\Romannum{2} Zeeman slowing case ($\Delta f \approx 1.6 \, \mathrm{GHz}$). This is because the laser only has to cover the lower hyperfine structure of $460 \, \mathrm{MHz}$ and additionally the changing Doppler shift. Therefore the spectral power density is higher and the slowing cuts deeper inside the longitudinal velocity distribution as the type-\Romannum{2} Zeeman slower, having a higher capture velocity of about $v_{\mathrm{cap}}=500 \, \mathrm{m \, s^{-1}}$. Still this benefit does not result in higher fluxes at low velocities.

 The flux of atoms below $35 \, \mathrm{m \, s^{-1}}$ achieved is about a factor of 20 less than for the type-\Romannum{2} Zeeman slower and correspondingly a factor of 33 lower than for the type-\Romannum{1} slower.
 Another disadvantage is the near resonant light of the slowing laser in the detection region which can disturb a following magneto-optical trap \cite{h_improved_nodate}. 


\section{Discussion}

Within this paper, we have demonstrated and compared three different beam slowing techniques. Two of these, white-light slowing and type-\Romannum{2} Zeeman slowing, are implementable for laser coolable molecules. 
As the magnitude of the slowing force for all three cases is on the same order we understood the observed differences to be the result of the efficiency with which this force is applied. 
Comparison of the experimental data with Monte Carlo simulations supported this idea.

We found type-\Romannum{2} Zeeman slowing to have the same continuous velocity compression characteristics as type-\Romannum{1} Zeeman slowing. In both schemes we achieve a well defined final velocity with a width limited by the resolution of our detection scheme.
The performance of the type-\Romannum{2} Zeeman slower is experimentally limited by the laser power of the frequency broadened repumper $\mathcal{L}_{\mathrm{rep}}$.
For the white-light slowing case the losses through transverse spreading are larger due to the unfavorable application of the slowing force.  Furthermore the slowing force has no sharp cutoff and the slowed atoms are spread out over a broad velocity range.
Compared to chirped light slowing, which was not implemented during this work and which is also applicable for laser coolable molecules, we expect the type-\Romannum{2} Zeeman slowing to be advantageous due to its continuous nature. The phase space acceptance of chirped light slowing is rather small and only parts of the comparatively long temporal pulse widths out of buffer gas sources can be slowed down.
Furthermore, particles reaching their final velocity determined by the end frequency of the chirp are spread out in space. Those disadvantages are partly compensated for by more relaxed power requirements allowing for larger slowing beam diameters in the experiment.
 The full potential of type-\Romannum{2} Zeeman slowing in comparison to chirped light might unfold in the future when realizing molecular experiments with continuous cold beam sources.

\end{document}